\documentclass[aps,10pt,prl,tightenlines,twocolumn,twoside,showpacs,superscriptaddress]{revtex4-1}
\usepackage{amsfonts}
\usepackage{float}
\usepackage{placeins}
\usepackage{graphicx}
\usepackage[hidelinks]{hyperref}
\usepackage{color,amsmath,amssymb,bm}
\usepackage{tensor}

\usepackage[normalem]{ulem}  % \sout{old text} for strikeout

\renewcommand{\sout}{\bgroup \color{red} \ULdepth=-.5ex \ULset}

\newcommand{\appropto}{\mathrel{\vcenter{
			\offinterlineskip\halign{\hfil$##$\cr
				\propto\cr\noalign{\kern2pt}\sim\cr\noalign{\kern-2pt}}}}}

\def\blfootnote{\xdef\@thefnmark{}\@footnotetext}
\newcommand{\beq}{\begin{equation}}
\newcommand{\eeq}{\end{equation}}
\newcommand{\bea}{\begin{eqnarray}}
\newcommand{\eea}{\end{eqnarray}}

\newcommand{\white}[1]{{\textcolor{white}{#1}}}
\newcommand\ddfrac[2]{\frac{\displaystyle #1}{\displaystyle #2}}

\def \be {\begin{equation} }
\def \ee {\end{equation}}
\def \bes {\begin{subequations} }
	\def \ees {\end{subequations}}

\def \a {\alpha}
\def \b {\beta}

\def \e {\varepsilon}
\def \g {\gamma}

\def \o {\omega}

\def \r {\rho}
\def \l {\lambda}

\def \s {\sigma}

\def \a {\alpha}
\def \b {\beta}

\def \e {\varepsilon}
\def \vp {\bm{p}}

\def \vx {\bm{x}}
\def \vy {\bm{y}}

\def \le {\left}
\def \ri {\right}
\def \<{\langle}
\def \>{\rangle}
\def \+{\dagger}

\def \[{\left[}
\def \]{\right]}

\def \pd {\partial}

\def \no {\nonumber}
\def \sA {{\cal A}}

\begin{document}

\title{Shear-induced spin polarization in heavy-ion collisions}
\author{Baochi Fu\footnote{fubaochi@pku.edu.cn}}
\affiliation{Department of Physics \& State Key Laboratory of Nuclear Physics and Technology, \\
   Peking University, Beijing 100871, China}
\affiliation{Collaborative Innovation Center of Quantum Matter, Beijing 100871, China}
\author{Shuai Y.\,F.~Liu\footnote{lshphy@gmail.com}}
\affiliation{Quark Matter Research Center, Institute of Modern Physics, Chinese Academy of Sciences, Lanzhou 730000, China}
\author{Longgang Pang\footnote{lgpang@mail.ccnu.edu.cn}}
\affiliation{Key Laboratory of Quark \& Lepton Physics (MOE) and Institute of Particle Physics, Central China Normal University, Wuhan 430079, China} 
\author{Huichao Song\footnote{huichaosong@pku.edu.cn}}
\affiliation{Department of Physics \& State Key Laboratory of Nuclear Physics and Technology, \\
   Peking University, Beijing 100871, China}
\affiliation{Collaborative Innovation Center of Quantum Matter, Beijing 100871, China}
\affiliation{Center for High Energy Physics, Peking University, Beijing 100871, China}
\author{Yi Yin\footnote{yiyin@impcas.ac.cn}}
\affiliation{Quark Matter Research Center, Institute of Modern Physics, Chinese Academy of Sciences, Lanzhou 730000, China}
\affiliation{University of Chinese Academy of Sciences, Beijing 100049, China}
\date{\today}

\begin{abstract}

We study the spin polarization generated by the hydrodynamic gradients. 
In addition to the widely studied thermal vorticity effects, we identify an undiscovered contribution from the fluid shear.
This shear-induced polarization (SIP) can be viewed as the fluid analog of strain-induced polarization observed in elastic and nematic materials. 
We obtain the explicit expression for SIP using the quantum kinetic equation and linear response theory.
Based on a realistic hydrodynamic model, we compute the differential spin polarization along both the beam direction $\hat{z}$ and the out-plane direction $\hat{y}$ in non-central heavy-ion collisions at $\sqrt{s_{NN}}=200$~GeV, including both SIP and thermal vorticity effects.
We find that SIP contribution always shows the same azimuthal angle dependence as experimental data and competes with thermal vorticity effects.
In the scenario that $\Lambda$ inherits and memorizes the spin polarization of strange quark, SIP wins the competition, and the resulting azimuthal angle dependent spin polarization $P_y$ and $P_z$ agrees qualitatively with the experimental data.
\end{abstract}

\maketitle

\textit{Introduction.---}The transport phenomena involving spin are instrumental in investigating quantum effects in many-body systems. 
For example, the generation of spin current can be employed to probe intriguing properties of quantum materials~\cite{SpinProbe}. 
In relativistic heavy-ion collisions, measuring spin polarization of hyperons has been proposed to explore the spin dynamics of quarks in the produced quark-gluon plasma (QGP) in heavy-ion collisions~\cite{Liang:2004ph}.
The observed $\Lambda$ spin polarization at RHIC and LHC experiments~\cite{STAR:2017ckg,Adam:2018ivw,Adam:2019srw,Niida:2018hfw}
 opens a new avenue to study the hot and dense nuclear matter\cite{
	%lett or other important
	Becattini:2013fla,Fang:2016vpj,Karpenko:2016jyx,Pang:2016igs,Becattini:2017gcx,Li:2017slc,Becattini:2018duy,Liu:2019krs,
	%review	
	Becattini:2020ngo,Huang:2020dtn,
	%spin hydro
	Florkowski:2017ruc,Hattori:2019lfp,Fukushima:2020ucl,Li:2020eon,Bhadury:2021oat,
	%spin trans
	Li:2019qkf,Wang:2020pej,Yang:2020hri,Weickgenannt:2020aaf,Weickgenannt:2021cuo,Li:2020vwh,
	%haronization
	Sheng:2020ghv,Wang:2021owk,
	%hydro ph
	Singh:2020rht,Wu:2019eyi,Fu:2020oxj,Florkowski:2019qdp,Florkowski:2019voj,
	%tran ph
	Sun:2017xhx,Sun:2018bjl,Xia:2018tes,Zhang:2019xya,
	%spindefine
	Speranza:2020ilk,
	%other interesting things
	Csernai:2018yok,Becattini:2020sww,Liu:2020dxg,liu2021spin}.

During the hydrodynamic evolution of the fireball created in heavy-ion collisions,
the sizable gradients of hydrodynamic fields, such as temperature and flow gradient, could give rise to spin polarization. 
One widely-studied effect is the spin polarization induced by thermal vorticity~\cite{Becattini:2013fla,Fang:2016vpj,Pang:2016igs}, which is a combination of temperature gradient and fluid vorticity.  
These theories based on the thermal vorticity effects successfully describe the global $\Lambda$ polarization measured in experiments~\cite{STAR:2017ckg,Adam:2018ivw}. 
In contrast, the predicted local (differential) $\Lambda$ spin polarization~\cite{Becattini:2017gcx,Xia:2018tes,Fu:2020oxj} differs qualitatively from experimental observations~\cite{Niida:2018hfw,Adam:2019srw}, 
see attempts to address such ``spin sign puzzle'' in Refs.~\cite{Liu:2019krs,Wu:2019eyi,Florkowski:2019voj}.

Nevertheless, vorticity and temperature gradient are not the only examples of hydrodynamic gradients.
In this letter, we identify the missing contribution to the spin polarization, namely, 
the effect of the shear stress tensor $\s^{\mu\nu}$, or fluid shear.
This shear-induced polarization (SIP) is not only allowed by symmetry but can also be derived explicitly based on quantum kinetic equation and the linear response theory, as we shall demonstrate later.
SIP can be viewed as the fluid analog of strain-induced polarization observed in elastic and nematic materials~\cite{Crooker_2005,2005PhRvL..95j7203M}.

Employing the data-calibrated hydrodynamic calculation~\cite{Fu:2020oxj}, we investigate spin polarization, including both SIP and thermal vorticity effects. We find in the scenario that $\Lambda$ inherits and memorizes the spin polarization of strange quark, SIP wins over the effects of thermal vorticity.
As such, the total spin polarization shows an azimuthal angle dependence qualitatively agrees with the experimental data.

\textit{Theory.---}%
Considering a system in the presence of slow varying flow velocity $u^{\mu}$ and temperature $T$, 
we look for gradient expansion of axial Wigner function $\sA^{\mu}$ which describes the phase space density of spin polarization of fermions.
% \red{the phase space density of the spin polarization of fermions including the SIP can be obtained through a gradient expansion of axial Wigner function $\sA^{\mu}$. }

\begin{figure*} [!t]
	\centering
	\includegraphics[trim={7cm 0 10cm 0}, clip, width=1.99\columnwidth]{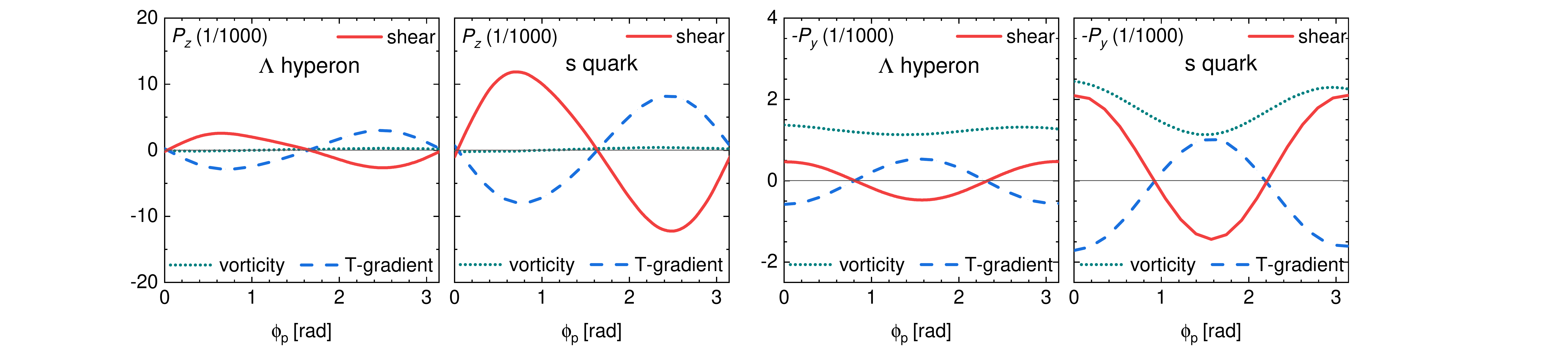} 	
	\caption{
  Spin polarization as a function of azimuthal angle $\phi_p$ along $z$ and $y$  directions, induced by hydrodynamic gradients for $\Lambda$ hyperon and strange quarks at the freezeout surface: colored curves show effects from vorticity, temperature gradient and shear stress tensor (i.e. the shear-induced polarization or SIP)
  , corresponding to the first, second and third terms in Eq.~\eqref{A-exp} respectively.
The effects of thermal vorticity are given by the sum of vorticity and temperature gradient effects and have been studied by many others.
SIP is the new effect studied in this letter and competes with the thermal vorticity effects. 
	}
	\label{fig_3term}
\end{figure*}

We will first study the simplest case that fermions are massless using the expression from chiral kinetic theory~\cite{Son:2012wh,Chen:2014cla,Chen:2015gta}:
%\begin{linenomath*}
\begin{align}
\label{A-CKT}
&\,\sA^\mu=\sum_{\l}\,\le( \lambda\, p^\mu\,f_{\l} +\frac{1}{2}\frac{\epsilon^{\mu\nu\alpha\rho}p_\nu u_\alpha \partial_{\rho} f_\lambda}{p\cdot u}\ri)\, , 
\end{align}
%\end{linenomath*}
where $\l=+/-$ accounts for right/left-handed chiral fermions respectively, $p_\mu$ is the momentum and $f_{\l}$ is the distribution function of particles.
The second term in Eq.~\eqref{A-CKT} is commonly referred as magnetic current MC term~\cite{Hattori:2019ahi}
which gives rise to various interesting transport phenomena such as off-equilibrium chiral magnetic effect~\cite{Kharzeev:2016sut}, spin Hall effect~\cite{Son:2012zy,Hattori:2019ahi,Liu:2020dxg}, and has been recently implemented into transport model for spin polarization~\cite{Liu:2019krs}.

Now, we replace $f_{\l}$ in Eq.~\eqref{A-CKT} with local equilibrium distribution function $n(\beta(\e_{0}-\Delta\e_{\l}))$ where the energy shift due to spin-vorticity coupling is given by $\Delta \e_{\l} =-(1/2)\lambda\o \cdot p/\e_{0}$ with $ \e_{0}=p\cdot u$ and $\o^\mu=(1/2)\epsilon^{\mu\nu\a\l}u_\nu\partial^{\perp}_\a u^{}_\l$ is the vorticity.
Here, $n(x)=1/(e^x+1) $ denotes Fermi-Dirac function. 
Then, we can expand Eq.~(\ref{A-CKT}) to first order in gradient as
%\begin{linenomath*}
\begin{align}
\label{A-exp-1}
\sA^\mu=\beta n_0(1-n_0)\Big[&%\red{-}
-\frac{1}{\e_{0}}p^\mu p^\nu\omega_\nu+\epsilon^{\mu\nu\a\r}u_{\nu} \,p_{\a}\beta^{-1}\pd_{\r}\beta
\no\\
-&\frac{1}{\e_{0}}\epsilon^{\mu\nu\a\rho}u_{\nu} \,p_{\r}p^\l\pd^{\perp}_{\a}u_\l\Big]\, .
\end{align}
%\end{linenomath*}
Here, we denote a generic vector $V^{\mu}$ projected along the transverse direction with respect to $u^{\mu}$ as $V^{\mu}_{\perp}=\Delta^{\mu\nu}V_{\nu}$ where $\Delta^{\mu\nu}=\eta^{\mu\nu}-u^\mu u^\nu$ and $\eta^{\mu\nu}=(1,-1,-1,-1)$ is the metric and $\epsilon^{0123}=1$.
We have also introduced the notation $n_0\equiv n(\beta \e_{0})$.
To proceed, we use 
$ \pd^\perp_{\a} u_\l=\epsilon_{\a\l\g\zeta}u^\g\o^{\zeta}+\pd^{\perp}_{(\a}u^{}_{\l)}$ to evaluate the last term in Eq.~\eqref{A-exp-1} where $\pd^{\perp}_{(\a}u^{}_{\l)}\equiv(\pd^{\perp}_{\a}u^{}_{\l}+\pd^{\perp}_{\l}u^{}_{\a})/2$.
Noting $ (\,-p_{\perp}\cdot \o)\, u^{\mu}+
\e_{0}\,\o^{\mu}=(1/2)\epsilon^{\mu\nu\a\l}p_\nu\partial^{\perp}_\a u_\l$ and $p^2=0$ for massless fermions \footnote{We have used $p^2=0$ to make the connection between the results from massless case and that of massive theory more transparent.},
we arrive at the desired expression
%\begin{linenomath*}
	\begin{align}
	\label{A-exp}
	\sA^{\mu}=& 
	\frac{1}{2}\b n_{0}(1- n_{0})\,\Big\{\epsilon^{\mu\nu\a\l}p_\nu\partial^{\perp}_\a u^{}_\l
	\no \\
	&+2\epsilon^{\mu\nu\a\l}  u_{\nu}p_{\a}\,[\b^{-1}(\pd_{\l} \beta)]\Big\}
	+\sA^{\mu}_{\text{SIP}}\,  
	\end{align}
%\end{linenomath*}
where the shear-induced spin polarization (SIP) as we advertised earlier is given by
%\begin{linenomath*}
\begin{align}
\label{SIP}
\sA^{\mu}_{\text{SIP}}&=-\beta n_{0}(1-n_{0}) \frac{1}{\e_{0}}\epsilon^{\mu\nu\a\rho}u_{\nu} \,p_{\r}p^{\l}\pd^{\perp}_{(\a}u^{}_{\l)}
\no\\
&= -\beta n_{0}(1-n_{0})\,\frac{p^{2}_{\perp}}{\e_{0}}\epsilon^{\mu\nu\a\r}u_{\nu}Q_{\a}^{\,\,\l}\s_{\rho\l}\,.
\end{align}
%\end{linenomath*}
From the first line to the second line in Eq.~\eqref{SIP}, we have used $ p^\mu=\e_{0} u^\mu +p^\mu_{\perp}$. 
The generalized quadrupole tensor and shear stress tensor are given by $ Q^{\mu\nu}\equiv- p^\mu_{\perp}p^\nu_{\perp}/p^2_\perp+\Delta^{\mu\nu}/3$ and $\s^{\mu\nu}=\pd^{(\mu}_{\perp}u^{\nu)}_{\white{\perp}}-\Delta^{\mu\nu}\partial\cdot u/3$ respectively. 
Note $\sA^{\mu}_{\text{SIP}}$ solely arises from the magnetic current term.

To extend our analysis to fermions with an arbitrary mass,
we consider the linear response theory (detailed in Ref~\cite{liu2021spin}) and find that the Eq.~\eqref{A-exp} applies equally to massless and massive fermions.
This means that the axial Wigner function $\sA^{\mu}$ only inexplicitly depends on the fermion mass through the mass dependence of $p^{\mu}$.
The key in the linear response analysis is to match the gradient expansion of $\sA^{\mu}$ to small frequency and wavevector behavior of the retarded correlator $G^{\a;\mu\nu}_{R}=\langle \bar{\psi}(t,\vx-\frac{\vy}{2})\,\g^{\a}\g^{5}\psi(t,\vx+\frac{\vy}{2})\, \hat{T}^{\mu\nu}(0,{\bm 0})\rangle\theta(t)$ where $\hat{T}^{\mu\nu}$ denotes the stress-energy tensor and $\psi$ is the fermionic field.
Using one loop calculation of $G^{\a;\mu\nu}_{R}$, we confirm Eq.~(\ref{A-exp}) for fermions with arbitrary mass.

For the convenience of the subsequent discussion, 
we combine the first two terms in ``\{\,\}" in Eq.~\eqref{A-exp} using the hydrodynamic equation $(u\cdot \pd) u_\a=-\beta^{-1}\pd^{\perp}_{\a}\beta+{\cal O}(\pd^{2})$ to obtain
%\begin{linenomath*}
\begin{align}
\label{A-final}
\sA^{\mu}
&=\frac{1}{2}n_0(1-n_0)\,
\epsilon^{\mu\nu\a\l}p_\nu\partial_\a (\beta u_\l)
+A^{\mu}_{\rm SIP}\,  
\end{align}
%\end{linenomath*}
where the first term reproduces the spin polarization induced by thermal vorticity, as was studied by many authors~\cite{Fang:2016vpj,Pang:2016igs,Becattini:2013fla}.
What is our new finding is the second term which describes the effects of shear stress tensor on spin polarization.

\begin{figure*}[!t]
	%\vspace{-1cm}
	\centering
	\includegraphics[trim={1cm 0 1cm 0}, clip, width=1.99\columnwidth]{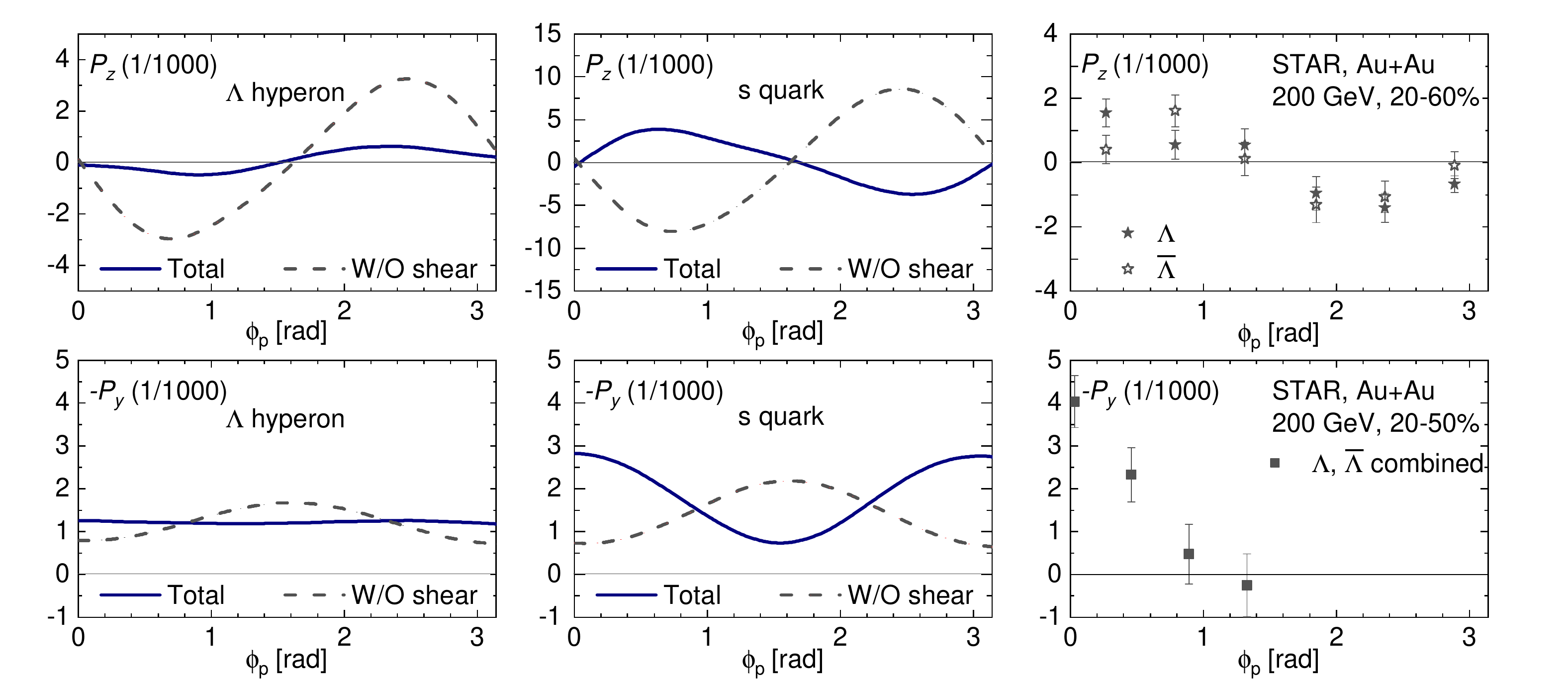}
	\caption{
		\label{fig:Ptotal}
		Spin polarization of $\Lambda$ hyperon (left) and strange quark (middle) along $ z $ (upper) and $ y $(lower) directions induced by the combined effects of shear stress tensor and thermal vorticity (solid curves) and by thermal vorticity effects only (dashed curves) on the freezeout surface. Right: the re-plotted experimental data in Ref~\cite{Niida:2018hfw,Adam:2019srw}.  The $P_z$ is converted using $P_z=\langle\cos \theta^*_p\rangle/[\alpha_H \langle(\cos \theta^*_p)^2\rangle]$ assuming zero error in denominator.
		The results for strange spin polarization illustrate the anticipated qualitative behavior in the "strange memory"scenario, i.e., the memory of strange quark polarization is preserved in the measured $\Lambda$ polarization.
	}
\end{figure*}

\textit{Two scenarios.---}To quantitatively predict the $\Lambda$ spin polarization due to hydrodynamic gradient effects, we need to hadronize the polarized strange quarks into Lambda hyperons followed by hadronic evolution towards the kinetic freezeout. 
However,  
the development of hadronization and transport models which consistently include the spin degrees of freedom is highly non-trivial, 
see Refs.~\cite{Sheng:2020ghv,Wang:2021owk} for recent progresses. 
In order not to introduce complicated model assumptions, we shall consider two widely assumed limiting scenarios~\cite{Karpenko:2016jyx,Pang:2016igs,Becattini:2017gcx,Fu:2020oxj,Liang:2004ph,Sun:2017xhx,Sun:2018bjl} and focus not on the magnitude but the qualitative features of the resulting azimuthal angle dependence of spin polarization.

In the first scenario,
namely the ``Lambda equilibrium'', 
we shall assume the spin relaxation rate is large enough so that $\Lambda$ hyperons immediately response to the presence of hydrodynamic gradients once $\Lambda$ are formed through hadronization.
In the second scenario, we consider the opposite limit that $\Lambda$ ``inherits'' the spin polarization from its constituent strange quark~\cite{Jennings:1990ui,Cohen:1991qi}, and the resulting spin polarization is frozen ever since the hadronization. 
This scenario will be referred to as the ``strange memory''.
In reality, $\Lambda$ spin polarization should evolve from the  ``strange memory" scenario towards that in the "Lambda equilibrium" scenario. 
Therefore comparing results from those two benchmark scenarios provides us qualitative guidance on what we anticipate to observe in heavy-ion collisions.

Guided by Refs~\cite{Becattini:2013fla,Fang:2016vpj},
we shall use the following freezeout prescription to connect axial Wigner function given by Eq.~\eqref{A-exp} to spin polarization vector $P^{\mu}$ on the freezeout hyper-surface $\Sigma_{\mu}$ :
%\begin{linenomath*}
	\begin{align}
	\label{Pol}
	P^{\mu}(\vp)=
	\ddfrac{
		\int d\Sigma^\a p_\a\, \sA^{\mu}(x,\vp;m)}{2 m\int d\Sigma^\a p_\a n(\beta \e_{0})}\, .
	\end{align}
%\end{linenomath*}
Here the factor of $2$ in the denominator counts two states of the spin-$1/2$ fermions. 
In the ``Lambda equilibrium" (``strange memory") scenario,
we shall compute $\Lambda$ (strange quark) spin polarization using Eq.~\eqref{Pol}. We use $m=1.116$~GeV for Lambda mass and the benchmark value for the strange quark mass $m=0.3$~GeV, respectively; the latter is between strange constituent and current mass. 
In principle, we should compute Eq.~\eqref{Pol} at the kinetic freezeout for ``Lambda equilibrium" scenario, but we have checked that the results of doing so are qualitatively similar to those calculated at chemical freezeout.
To simplify the comparison, 
we shall show spin polarization vector computed at chemical freezeout in both scenarios.

\textit{Results--}
In this letter, we implement 3+1-d hydrodynamics MUSIC~\cite{Schenke:2010nt,Schenke:2010rr,Schenke:2011bn} with AMPT initial conditions~\cite{Lin:2004en,Pang:2012he, Xu:2016hmp} to generate the freeze-out surface and associated temperature and flow velocity profiles for the spin polarization calculation described by Eq.(3). 
Unless noted otherwise, we use the same inputs and parameter sets as used in previous paper~\cite{Fu:2020oxj} where the ``Lambda equilibrium'' scenario without SIP has been studied. 
Such hydrodynamic calculations have been well calibrated to fit the $dN_{\text{ch}}/dy$, $p_T$ spectra and $v_2(p_T)$ of pions and protons in Au-Au collisions at $\sqrt{s_{NN}}=200$~GeV.
\footnote{
We have also performed an independent calculation using CLVisc~\cite{Pang:2016igs} hydrodynamics framework and produce all the key features shown in the main text. 
}.
In what follows, we will focus on 
the spin polarization vector along the beam direction, $P_z(\phi_p)$, and along the out-plane direction, $P_y(\phi_p)$,
as a function of the azimuthal angle $\phi_p$.

In Fig.~\ref{fig_3term} and Fig.~\ref{fig:Ptotal}, we plot the differential spin polarization vector along the beam direction, $P_{z}(\phi_{p})$, and along the out-plane direction, $P_{y}(\phi_{p})$, in the particle rest frame.
They include the separate and/or combined effects from SIP, fluid vorticity and temperature gradient. 
Although the thermal vorticity contribution has been studied extensively,
this is the first time that the signature of SIP has been investigated in heavy-ion collisions. 
All curves contributing to $P_z$ and $(-P_y)$ can be parametrized approximately as ``$b_z\sin(2\phi_p) $" and ``$a+b_y\cos(2\phi_p)$" respectively. 
For the transparency of comparing with the qualitative features of the experimental results shown in Fig.~\ref{fig:Ptotal} ,
we shall refer a contribution as the "same" sign (``opposite" sign) when $b_z, b_y>0$ ($b_z, b_y<0$).

In Fig.~\ref{fig_3term}, we plot spin polarization induced by shear stress tensor (SIP), vorticity (VoIP) and temperature gradient (TIP) (c.f.~Eq.~(\ref{A-exp})) in both ``Lambda equilibrium" and ``strange memory" scenarios. 
We observe that the contribution from SIP and TIP to global polarization is insignificant.
However, 
the azimuthal angle dependence of $P_z$ and $P_{y}$ arises mostly from SIP and TIP. 
As a marked qualitative feature, SIP always leads to the ``same" sign contribution to spin polarization in both $z$ and $y$ directions.
On the other hand, the effects of thermal vorticity on the azimuthal angle dependence of spin polarization are dominated by TIP, which shows the ``opposite" sign behavior. 
That thermal vorticity leads to the ``opposite'' sign, as seen in many early studies~\cite{Becattini:2017gcx,Xia:2018tes,Fu:2020oxj}, is sometimes referred to as ``spin sign puzzle''.

Given SIP contribution is comparable to TIP contribution in magnitude but is qualitatively different in "sign", 
the competition between SIP and TIP will eventually determine the azimuthal angle dependence of the total spin polarization.
This competition is best seen in Fig.~\ref{fig:Ptotal}, where the total spin polarization as a function of $\phi_p$ are drastically different with and without SIP. 
In ``Lambda equilibrium" scenario, though, 
TIP wins the competition and the total polarization is the ``opposite" sign even in the presence of SIP.
However, 
SIP becomes more important when the mass of spin carrier becomes smaller.
This can be easily understood from Eqs.~\eqref{SIP},\eqref{A-final} that the factor $|p^2_{\perp}|/\e^{2}_{0}$ (the square of typical velocity of fermions) is larger when the spin carrier is lighter. 
Indeed, in the ``strange memory" scenario, SIP prevails over TIP  in both $P_{z}$ and $P_{y}$. 
Seeing this, we should not be surprised to find that the total spin polarization shows the ``same" sign.

\begin{figure} [t]
	\centering
    \includegraphics[trim={1cm, 2.5cm, 1cm, 1cm},clip, width=0.65\columnwidth]{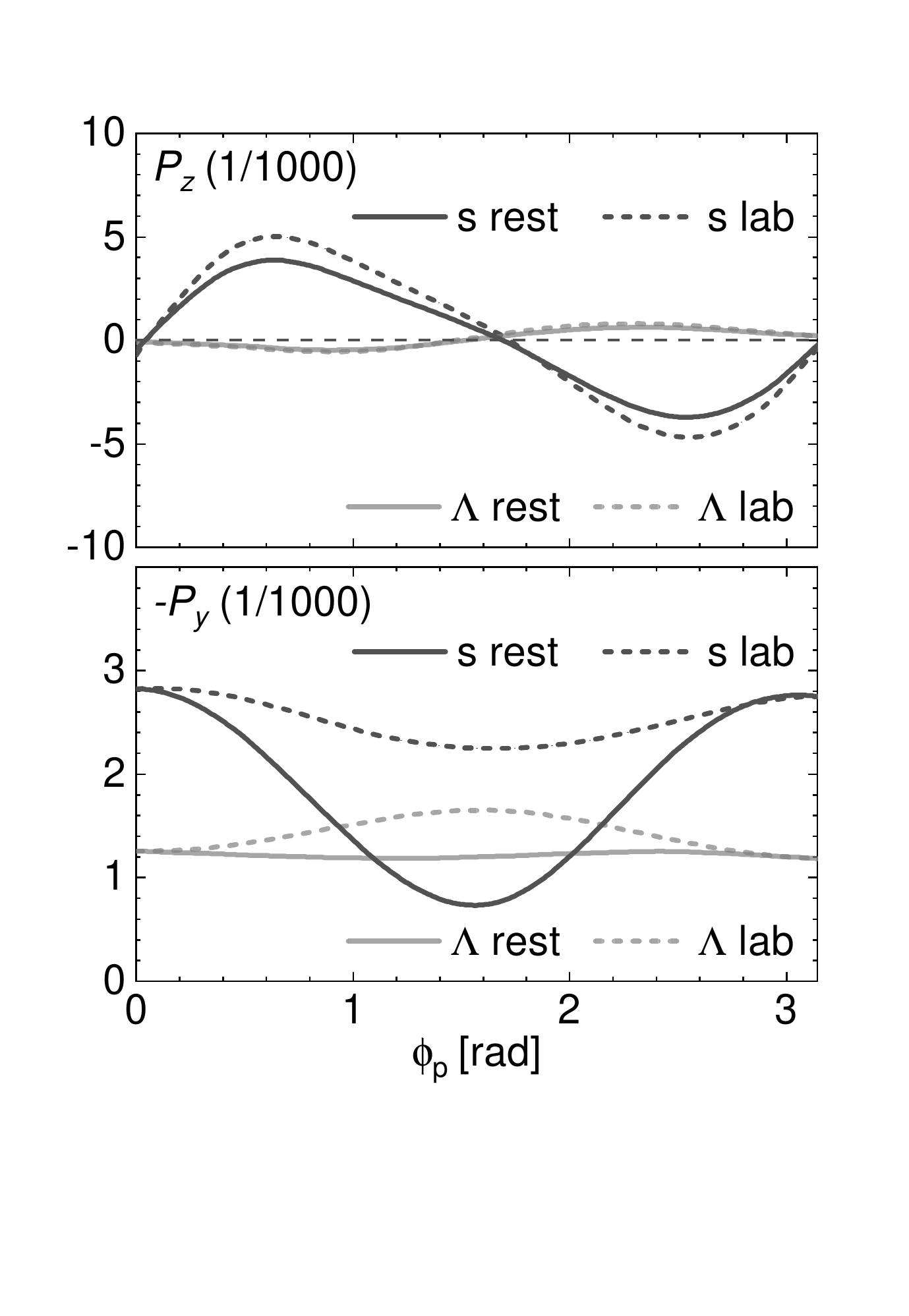}
	\caption{The comparison between strange quark and $\Lambda$ hyperon spin polarization in lab and particle rest frame.}
	\label{fig_frame}
\end{figure}

To complement Figs.~\ref{fig_3term}~\ref{fig:Ptotal},
we compare our results in the particle rest frame as shown above with those in the lab frame in Fig.~\ref{fig_frame}.
This comparison illustrates the sensitivity of spin polarization to the choice of the reference frame. 
 We notice that $ P_y$ is more sensitive to the change of the reference frame than $P_z$; see also Ref~\cite{Liu:2019krs} for related discussion.

To investigate the generality of the results reported above, 
we repeat our calculations by systematically varying inputs for the present hydrodynamic model, including initial conditions, freezeout temperature, and the parametrization of the shear viscosity, bulk viscosity and equation of state (EoS), see our upcoming publication for more details. 
In all cases that we have explored, 
SIP always induces the ``same sign" contribution whereas TIP contribution is the ``opposite sign". 
In ``Lambda equilibrium" scenario, the total spin polarization is always "opposite" sign. 
This should be contrasted with the results seen in the ``strange memory" scenario. 
Except for some specific cases when a unusually large $T$-gradient is caused by EoSs much harder than the lattice EoS or a sharply-peaked bulk viscosity around the freezeout region, 
the azimuthal angle dependence of spin polarization is generically dominated by SIP and shows the qualitative agreement with the data.

\textit{Summary and discussion.---}In this letter, we investigate the shear-induced spin polarization (SIP), which is an undiscovered mechanism for spin polarization generation. 
We study the effects of SIP and thermal vorticity on the azimuthal angle dependence of $\Lambda$ spin polarization in heavy-ion collisions at $\sqrt{s_{NN}}=200$~GeV based on a data-calibrated hydrodynamic model. 
Our main results, summarized in Fig.~\ref{fig_3term} and Fig.~\ref{fig:Ptotal},
clearly demonstrate that
SIP gives rise to distinctive new features in differential spin polarization and is indispensable to analyze the effects induced by the hydrodynamic gradient on the measured spin polarization.
By contrasting the results in the ``Lambda equilibrium" scenario where the strange quark's memory is completely forgotten and those in the ``strange memory" scenario, 
it is tempting to conclude that the presence of both SIP and the memory of strange quarks is required for the azimuthal angle dependence of spin polarization to agree with data qualitatively.

For future quantitative studies, 
it is important to understand how to form a polarized $\Lambda$ from the polarized quarks and the subsequent evolution in the hadronic stage~\cite{	Li:2019qkf,Wang:2020pej,Yang:2020hri,Weickgenannt:2020aaf,Weickgenannt:2021cuo,Li:2020vwh,Sheng:2020ghv,Wang:2021owk}. 
Future studies should also investigate the evolution of quark spin polarization in QGP~\cite{Florkowski:2017ruc,Hattori:2019lfp,Fukushima:2020ucl,Li:2020eon,Bhadury:2021oat}. 
We limit our calculations to high-energy heavy-ion collisions. 
SIP should be present in collisions at the beam scan energies at RHIC and could potentially be employed to explore the properties of QCD matter at finite baryon density.

\acknowledgments
We thank helpful discussions with Shanshan Cao, Xu Cao, Hengtong Ding, Fei Gao, Feng Li, Yanf and Yifeng Sun.
This work was supported in part by the NSFC under grant No.~12075007 and No.~11675004 (BF and HS) and by No.~11861131009 and No.~12075098 (LP) as well as by
the Strategic Priority Research Program of Chinese Academy of Sciences, Grant No. XDB34000000 (SL and YY).
We acknowledge the extensive computing resources provided by the Supercomputing Center of Chinese Academy of Science (SCCAS), Tianhe-1A from the National Supercomputing Center in Tianjin, China and the High-performance Computing Platform of Peking University as well as Nuclear Science Computer Center at CCNU (NSC3), China.

\bibliography{refcnew}
\end{document}